\documentclass[a4paper,twocolumn]{esapub2005} 
\usepackage{times}

\usepackage{natbib}

\usepackage{graphicx}

\title{Following the Colour of the Low Mass X-ray Binary 4U 1820--30 with {\it INTEGRAL\/}}
\author{A. Tarana}
\affil{IASF-INAF Rome, Italy; Universit\`a Tor Vergata di Roma, Italy}
\author{A. Bazzano}
\author{P. Ubertini}
\affil{IASF-INAF Rome, Italy}
\author{A. A. Zdziarski}
\affil{Copernicus Astronomical Center, Warsaw, Poland}
\author{M. Federici}
\affil{IASF-INAF Rome, Italy}

\newcommand{\source}{4U 1820--30}

\begin{document}
\keywords{X-ray binaries; neutron star; individual: 4U 1820--30}
\maketitle

\begin{abstract}

The 4--200 keV spectral and temporal behaviour of the low mass X-ray binary 4U
1820--30 has been studied with {\it INTEGRAL\/} during 2003-2005. This source as been observed in both the soft (banana) and hard (island) spectral states. A high energy tail above 50 keV in the hard state has been revealed for the first time. This places the source in the category of X-ray bursters showing high-energy emission. The tail can be modeled as a soft power law component, with the photon index of $\simeq$ 2.4, on top of thermal Comptonization emission from a plasma with the electron temperature of $kT_{\rm e} \simeq$6 keV and optical depth of  $\tau \simeq$ 4. Alternatively, but at a lower goodness of the fit, the hard-state broad band spectrum can be accounted for by emission from a hybrid, thermal-nonthermal, plasma. During the observations, the source spent most of the time in the soft state, as previously reported and the $\ge$4 keV spectra can be represented by thermal Comptonization with $kT_{\rm e} \simeq$3 keV  and $\tau \simeq$6--7.

\end{abstract}

\section{Introduction and data analisys}

4U 1820$-$30 is a low mass X-ray binary seen at 0.66$^{\prime\prime}$ from the centre of the globular cluster NGC 6624. It was the first identified source of type-I X-ray bursts \citep{grindlay}. Its distance has been estimated as $d=5.8$--7.6 kpc (e.g., \citep{Kuu}; \citep{shapo}).

The binary consists of a He white dwarf of the mass of (0.06--$0.08)M_{\odot}$ \citep{rappa} and a neutron star, with the mass estimated by \citep{shapo} as $\sim\! 1.3M_{\odot}$, orbitting at the short period of 11.4 m \citep{Stella}. In X-rays, \source\ is classified as an atoll \citep{hasinger}. However, its flux variation between the soft (banana) state to the hard (island) state are quasiperiodic at $\sim$170 d (\citep{priedho}, \citep{simon}, \citep{wen}), which has been proposed to be due to tidal effects of a more remote third star (\citep{cho}, \citep{Zdz06}). X-ray bursts occur only at low flux levels (e.g.,\citep{cho}), and the frequencies of its kHz QPOs are correlated with the flux \citep{zha}. X-ray spectra of the source were fitted by thermal Comptonization with or without an additional blackbody (\citep{bloser}; \citep{migliari}).

We report the first detection of X-ray emission above 50 keV from \source, in a hard spectral state. We study the source in the 4--200 keV energy range with the JEM-X (4--35 keV) \citep{lund} and IBIS (15 keV--1 MeV) \citep{uber} instruments on board the {\it INTEGRAL\/} satellite \citep{wink}, within the Galactic Centre Deep Exposure (GCDE) programme \citep{wink2}.



We have analised all the available data during which the source was within the IBIS/ISGRI (15 keV--1 MeV) and JEM-X detectors fully coded FOV ($9^{\circ} \times$9$^{\circ}$ and $4.8^{\circ}\times 4.8^{\circ}$, respecitvely) so the flux evaluation is not affected  by calibration uncertainties in the off-axis response. This yields 308 and 51 Science Windows, SCWs (pontings lasting about 2000 s) for the IBIS and JEM-X, respetively, during the revolutions 50--363. The data were extracted with the Off-Line Scientific Analysis (OSA) \citep{gold} v.\ 5.1 software released by the {\it INTEGRAL\/} Science Data Centre \citep{courvosier}. The spectral analysis was done with the XSPEC package v.\ 11.3.

We combine three methods of analysis: {\it Temporal}, using the light curves in different energy bands; {\it Photometric}, with the Hardness--Instensity Diagram; and {\it Spectral}, by spectral fitting.

\begin{figure}
\centering
\includegraphics[height=8.0cm,angle=90]{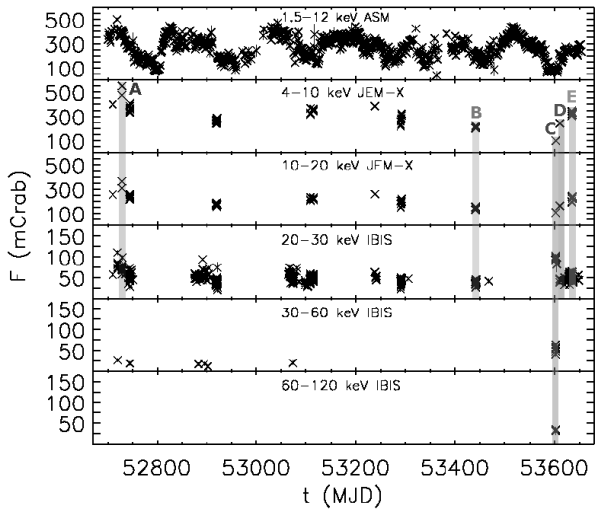}
\caption{The 2003--2005 light curves of \source, with the detector count rate in each band given with respect to the corresponding Crab count rate. The panels are marked with the energy range and detector, and present the {\it INTEGRAL\/} SCW data points except for the top panel, which gives the {\it RXTE}/ASM 1-day averages. The lines mark the data sets (A, B, C, D, E) used for joint IBIS and JEM-X spectral fits.\label{fig1}}
\end{figure}

\section{Light curves and hardness--intensity Diagrams}

Fig.\ \ref{fig1} shows the light curves of the monitoring period (2003 March 12 to 2005 October 5) from soft to hard X-rays, in the 4--10, 10--20 keV bands from the JEM-X and 20--30, 30--60, 60--120 keV with the IBIS/ISGRI, as well as the corresponding {\it RXTE}/ASM\footnote{http://xte.mit.edu/ASM$\_$lc.html.} light curve. The $\sim$170-d quasiperiodic variability can be seen in the ASM data. 

The {\it INTEGRAL\/} light curves are marked with vertical lines to denote the epochs (A, B, C, D, E) during which we have performed detailed spectral analysis. The epoch A corresponds to the joint IBIS and JEM-X count rate maximum ($\sim$530 mCrab at 4--10 keV). The epoch C corresponds to the minimum of the 4--10 keV count rate ($\sim$100 mCrab), but to high count rates from 20 to 60 keV, implying a spectral hardening across the 4--60 keV band. We also studied three epochs before (B) and after (D, E) that event.

\begin{figure}[ht]
\centering
\includegraphics[height=3.95cm,angle=90]{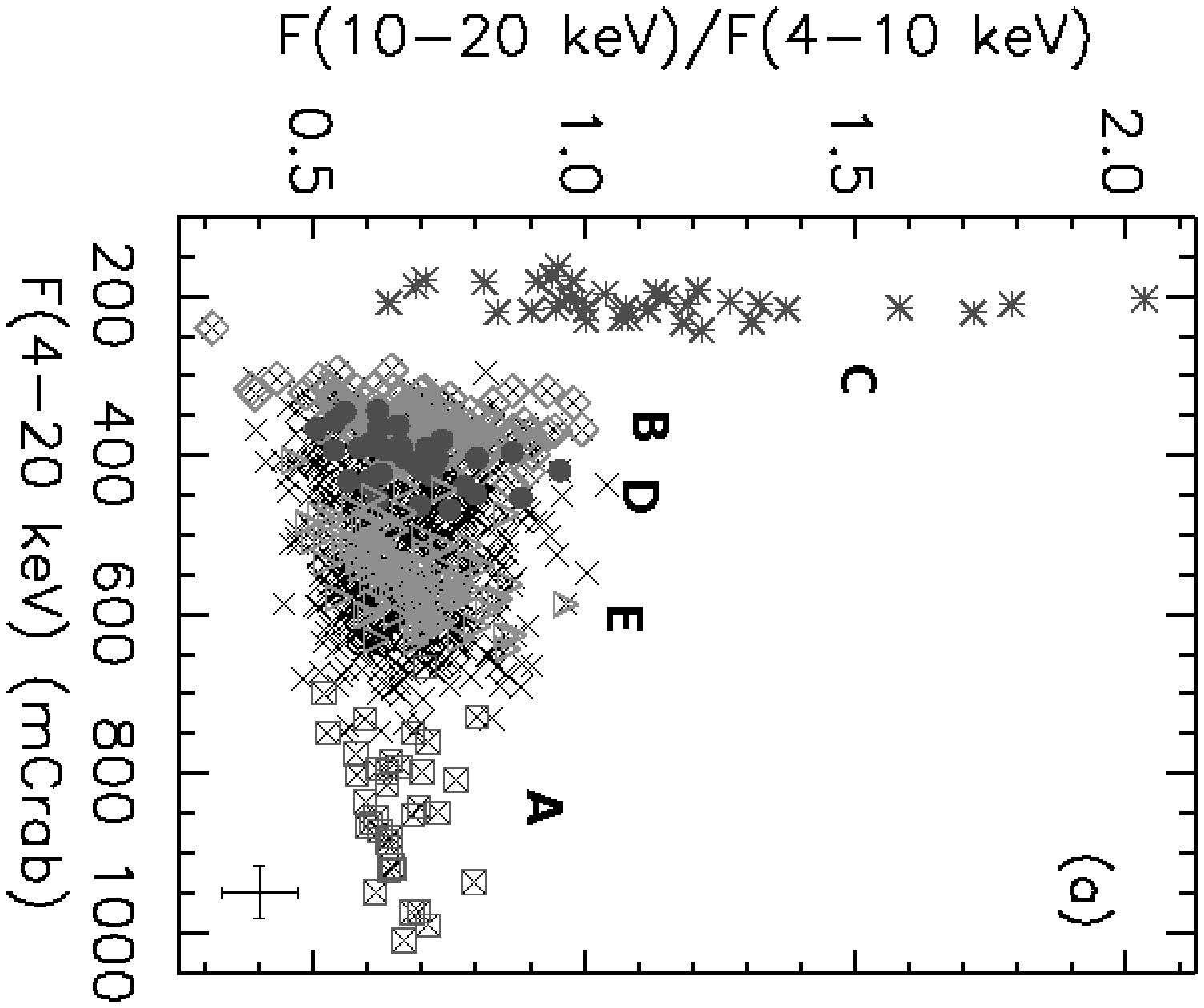}
\includegraphics[height=3.95cm,angle=90]{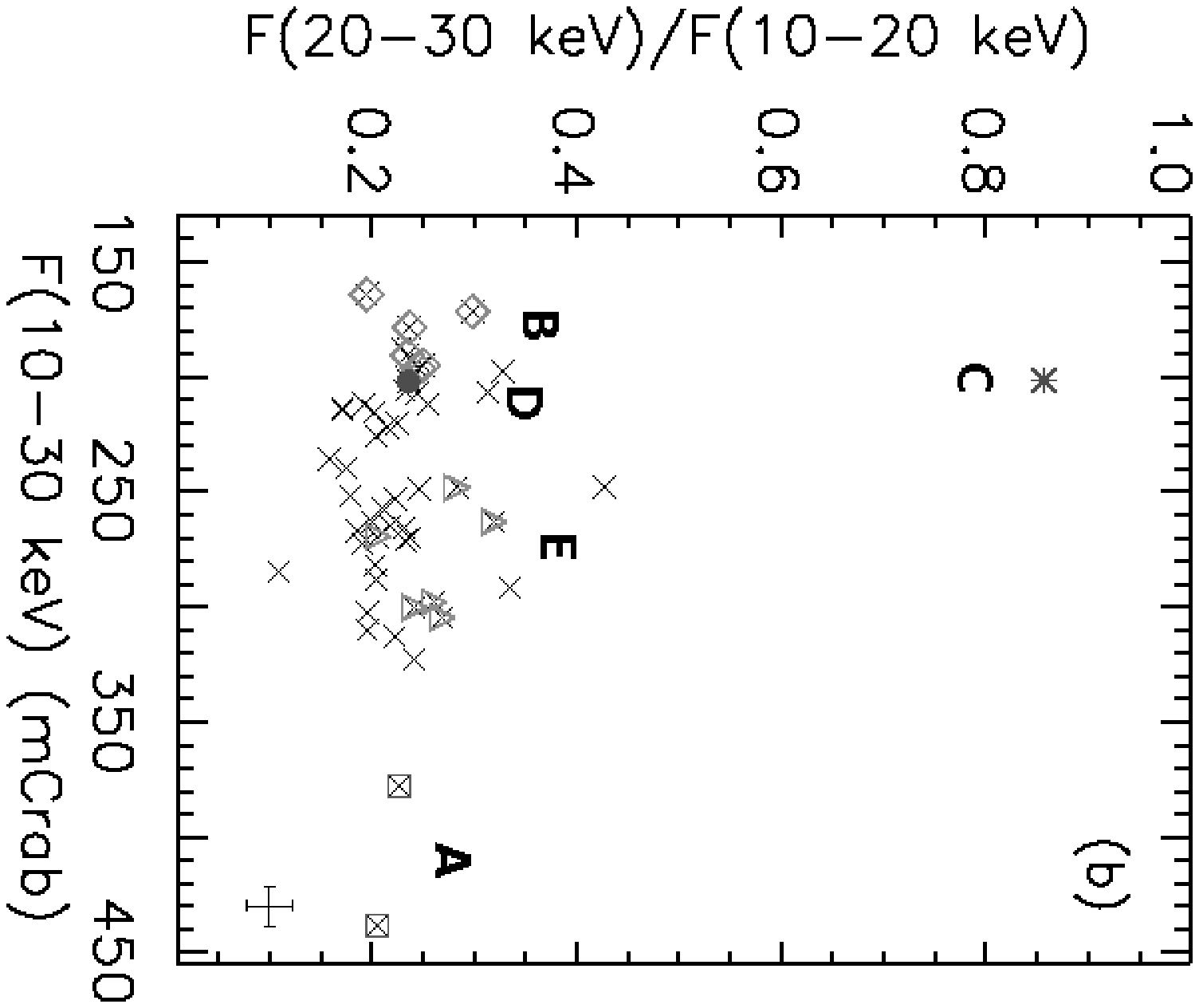}
\caption{(a) Hard color--intensity diagram; each point corresponds to 100 s. (b) Very hard color--intensity diagram; each point corresponds to a SCW. The letters identify the A--E data sets, see Fig.\ \ref{fig1}. The typical error bars are shown at the bottom right corbers. 
\label{fig2}}
\end{figure}

Fig. \ref{fig2} shows hardness--intensity diagrams for the data from the (a) JEM-X and (b) IBIS/JEM-X (for simultaneous pointings and when possible because of different FOV). Fig.\ \ref{fig2}b, for 10--30 keV, allows us to study spectral variability of the source at an energy band above that usually studied. In Figs. \ref{fig2}a, b, we can see the banana state, forming nearly horizontal bands across the A, B, D, E observations. Thus, the source evolution is mostly in the flux, with the 4--20 keV spectral shape close to constant. The upper part of the C points (asterisks) in Fig.\ \ref{fig2}a corresponds to the island state. We see that now the 4--20 keV flux is close to constant but the hardness shows large changes. This indicates spectral pivoting somewhere in the middle of that energy band. In Fig. \ref{fig2}b, we see similar behaviour, with the horizontal banana state. We have here only one pointing for the island state, with a large value of the hard color. These findings are confirmed below by spectral analysis. 

\section{Spectral analisys}

We fit the data with a number of models and their combinations, namely, thermal and hybrid Comptonization, blackbody, disk blackbody and a power law.
\begin{figure}[ht]
\centering
\includegraphics[height=7.4cm,angle=-90]{banana.ps}
\includegraphics[height=7.4cm,angle=-90]{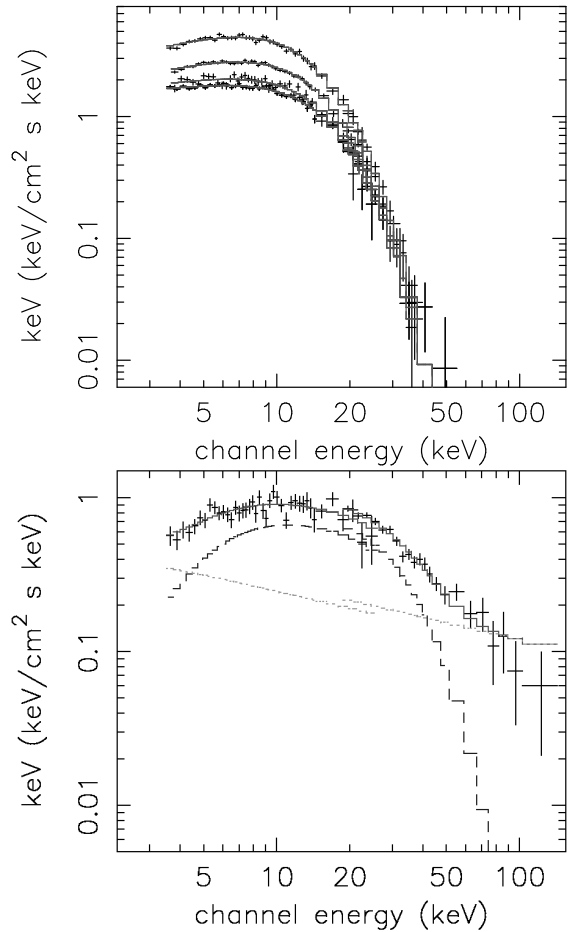}
\caption{Data, models of the soft spectra A, B, D, E (top) and the hard spectrum C (bottom), fitted by thermal Comptonization without and with, respectively, a power law component.
\label{SPE_ldata}}
\end{figure}
\begin{table*}
\begin{center}
\caption{Spectral fitting results for the JEM-X and IBIS broad-band spectra. The model is CompTT for A, B, D, E (banana states), and a CompTT + a power law for C (island state).} \vspace{1em}
    \renewcommand{\arraystretch}{1.2}
\begin{tabular}{l|ccccc}
\hline
parameters & A  & B  &  C   &  D  & E \\
\hline
$kT_{0}$ (keV) (frozen) & 0.2                    & 0.4                    & 1.5                    & 0.3                    & 0.2                     \\
$kT_{\rm e}$ (keV)      & 2.71$^{+0.08}_{-0.07}$ & 3.12$_{-0.10}^{+0.10}$ & 6.10$^{+0.77}_{-0.62}$ & 2.90$_{-0.18}^{+0.20}$ & 2.83$_{-0.06}^{+0.07}$  \\
$\tau$                  & 7.04$^{+0.32}_{-0.29}$ & 5.85$_{-0.23}^{+0.24}$ & 3.89$_{-0.65}^{+0.79}$ & 6.38$_{-0.62}^{+0.72}$ & 6.71$_{-0.22}^{+0.24}$  \\
norm$_{\rm CompTT}$     & 2.34$^{+0.21}_{-0.19}$ & 0.58$^{+0.06}_{-0.06}$ & 3.79$^{+0.68}_{-0.59}$ $\times 10^{-2}$ & 1.06$_{-0.18}^{+0.21}$&1.57$_{-0.11}^{+0.12}$ \\
\hline
$\Gamma$                & $-$                    & $-$                    & 2.35 $_{-0.11}^{+0.10}$ & $-$                   & $-$  \\
norm$_{pl}$             & $-$                    &$-$                     & 0.60$_{-0.24}^{+0.22}$  & $-$                   & $-$  \\
\hline
$\chi_\nu^2$(d.o.f)     &1.13(44)                & 0.96(46)               &1.0(57)                  & 1.07(37)              & 1.14(43)\\
\hline
$F_{\rm 4-20 keV}$ (erg s$^{-1}$ cm$^{-2}$)  & 9.1$\times 10^{-9}$ & 3.8$\times 10^{-9}$ & 2.1$\times 10^{-9}$  & 4.2$\times 10^{-9}$  & 5.8$\times 10^{-9}$\\
$F_{\rm 20-60 keV}$ (erg s$^{-1}$ cm$^{-2}$) & 3.3$\times 10^{-10}$& 1.8$\times 10^{-10}$& 7.4$\times 10^{-10}$ & 2.2$\times 10^{-10}$ & 2.5$\times 10^{-10}$  \\
$F_{\rm 60-120 keV}$ (erg s$^{-1}$ cm$^{-2}$)& $-$                 & $-$                 & 1.5$\times 10^{-10}$ & $-$ & $-$ \\
\hline
\end{tabular}
\label{tab:tabspetot}
\end{center}
\end{table*}

First, we find a thermal Comptonization model, CompTT \citep{Tit}, to sufficiently model the spectra of all of the banana states (A, B, D, E).
We obtain the electron temperature, $kT_{\rm e}\simeq 2.7$--3.1 keV, and the optical depth $\tau\simeq 6$--7, given in Table \ref{tab:tabspetot}. This is consistent with the color-intensity diagram, where we found only flux variation (most likely proportional to the accretion rate). The temperatures of the seed photons (Table \ref{tab:tabspetot}) have been frozen at their best-fit value, given that those photons are entirely below the fitted band. The data and model for the banana spectra
are shown in top of Fig.\ \ref{SPE_ldata}, which have the highest unabsorbed bolometric luminosity of $7.7\times 10^{37}$ erg s$^{-1}$. An addition of a blackbody component \citep{mitsuda} does not significantly improve the fits, 
and thus the confidence ranges of the temperature, $kT_{\rm bb}$, are not constrained. Still, we find $kT_{\rm bb}\simeq 1.5$--2.4 keV at the best fits, in agreement with the previous study \citep{bloser}.

The spectrum of the island state (C) shows an evident excess above 50 keV that cannot be fitted by thermal Comptonization only.
An addition of a soft ($\Gamma\simeq 2.4$) power law improves the fit significantly ($\chi^2_\nu=1.0$)
with the chance probability from the F-test
of $6.3\times 10^{-7}$. The bottom of Fig.\ \ref{SPE_ldata} shows the C-spectrum best fit with the comptonization and power law component (dotted line). The parameters of the thermal plasma are different from those for the banana state, with a higher $kT_{\rm e}$ ($\simeq 6$ keV), and a lower $\tau$ ($\simeq 4$), see Table \ref{tab:tabspetot}. By adding a disk blackbody component to the thermal Compton model, we have also obtained a good fit, with $\chi_\nu^2 =1.0$. However, the required inner temperature, $kT_{\rm in}\simeq 5.4$ keV, is clearly unphysical. Addition of a single blackbody gave a worse fit, as well as an unphysically high temperature, $\simeq 4$ keV.

We have also fitted the island-state data with the Comptonization model EQPAIR (\citep{coppi}, \citep{gie99}). We have first tried to account for the high-energy tail by adding nonthermal electrons to the thermal plasma, as done to account for the soft-state spectra of both black hole \citep{gie99} and neutron-star \citep{farin} binaries. We have obtained a moderately good fit, $\chi_\nu^2 =1.15$, with a model consisting of a thermal plasma with $kT_{\rm e}\simeq 8$ keV and $\tau\simeq 4$, and a high-energy tail resulting from electron acceleration at a power-law index of 2.5, and with an approximate equipartition between the powers in the heating of the thermal plasma and the acceleration. The bolometric luminosity of this model is $1.8\times 10^{37}$ erg s$^{-1}$. 
With the EQPAIR model used to model thermal Comptonization (instead of CompTT), i.e., with the null fraction of the power in acceleration, addition of a disk blackody component resulted in a fit improvement and a likely value of the inner temperature, $kT_{\rm in}\simeq 0.9$ keV. A further significant improvement resulted from taking into account Compton reflection \citep{refl}, detected in atolls (e.g., \citep{done}). The resulting model has $\chi_\nu^2 =1.09$, $kT_{\rm e}\simeq 22$ keV, $\tau\simeq 1.5$, similar to the parameters found in the atoll source 4U 1608--52 \citep{done}. Furthermore, we also used the COMPPS model for modelling the hard state data, see fit details results in \citep{tarana}. We note, however, that these last models are significantly more complex than the one with CompTT and a power law presented above, and still yields significant residuals at energies $>80$ keV. 

\section{Discussion and conclusion}

We have presented the first detection of hard tail, above 50 keV, in the island state of the atoll \source, using {\it INTEGRAL}/IBIS. Up to now, \source\ has not been included in the sample of the high-energy emitting bursters \citep{bazzano} and indeed neither {\it BeppoSAX}, {\it CGRO}/BATSE  nor {\it RXTE\/} detected this source above 50 keV (\citep{bloser} and references therein). The previous lack of detection might have been due to either the source spending most of the time in the soft state, the weakness of the emission above 50 keV in the island state, or the tail appearing in that state only occasionally, due to some process like jet formation. 

In our island-state observation, the relative contribution of the energy flux $>60$ keV is $\sim$10\%. The ratio of the 4--20 keV flux to the 20--120 keV one is  $\sim$2, whereas it it is un the range 20--27 for the soft (banana) states. We point out that we have detected the emission above 60 keV when the 1--20 keV luminosity was $\simeq 1.4 \times 10^{37}$ erg s$^{-1}$, i.e., somewhat below the critical value at which the X-ray binaries become hard emitters as proposed by Barret \citep{barret1996}, \citep{barret2000}. 

While the origin of the thermal Comptonization component appears well understood as emission from hot electrons that upscatter soft photons coming from the accretion disk and the neutron star surface, the origin of the high-energy tail is less clear. The radio emission detected from \source\ \citep{migliari} (even if during soft state) suggests the presence of a jet, which may also produce X-rays. 
But in this case, a more likely origin of the power law tail appears to be emission of nonthermal electrons in a hybrid (i.e., containing also thermal electrons) plasma, as discussed in Section 3.

Nonthermal tails have been commonly seen in soft states of black-hole binaries (e.g., Cyg X-1, \citep{gie99}) and neutron-star Z-sources (e.g., GX~17+2 \citep{disalvo}, \citep{farin}, Sco~X-1 \citep{damico}, GX~349+2 \citep{disalvo_b}, GX~5--1 \citep{asai}), while they are difficult to detect in atoll sources (e.g., 4U 0614+091 \citep{pira}). On the other hand, nonthermal tails have been claimed in the hard state of some black-hole binaries (Cyg X-1 \citep{McConnell} and GX 339--4 \citep{ward}).

\section*{Acknowledgments}

This research has made use of data obtained through the {\it INTEGRAL\/} Science Data Centre, Versoix, Switzerland. The authors thanks G. De Cesare and L. Natalucci for the scientific and data analysis support. This work has been supported by Italian Space Agency through the grant I/R/046/04. AAZ has been supported by the Polish grants 1P03D01827, 1P03D01128 and 4T12E04727. The authors thanks to  Catia Spalletta for the logistic support.






\end{document}